\documentclass[aps,prl,twocolumn,10 pt,groupaddress,floats,showpacs]{revtex4}
\usepackage[dvips]{graphicx}
\usepackage{amsmath}

\begin{document}

\title{The influence of twin boundaries on the Flux Line Lattice structure in YBaCuO: a study by Small Angle Neutron Scattering. }
\author{Ch. Simon, A. Pautrat, G. Poulain, C. Goupil, C. Leblond-Harnois}
\affiliation{Laboratoire CRISMAT, UMR 6508 du CNRS et de l'ENSI-Caen, 14050 CAEN,
France.}
\author{ X. Chaud}
\affiliation{CNRS/CRETA 25, avenue des martyrs BP 166 38042 Grenoble Cedex 09, France.}
\author{ A.\ Br\^{u}let}
\affiliation{Laboratoire L\'{e}on Brillouin, CEA-CNRS\\ CEN Saclay, 91191 Gif/Yvette,
France.}

\begin{abstract}
The influence of Twin Boundaries (TB) on the Flux Line Lattice(FLL) structure was
investigated by Small Angle Neutron Scattering (SANS). YBaCuO single crystals possessing
different TB densities were studied. The SANS experiments show that the TB strongly
modify the structure of the FLL. The flux lines meander as soon as the magnetic field
makes an angle with the TB direction. According to the value of this angle but also to
the ratio of the flux lines density over the TB density, one observes that the FLL
exhibits two different unit cells in the plane perpendicular to the magnetic field. One
is the classical hexagonal and anisotropic cell while the other is affected by an
additional deformation induced by the TB. We discuss a possible relation between this
deformation and the increase of the critical current usually observed in heavily twinned
samples.
\end{abstract}

 \maketitle

\newpage

\section{Introduction}

The role of the twin boundaries (TB) on the Flux Lines Lattice (FLL) behavior in YBaCuO
was the subject of an intense activity ten years ago. Nevertheless, most of the results
from the different experiments remain questionable due to the difficulty in isolating the
effective role of TB. The growth of untwinned single crystals has recently allowed to
clarify this point, especially from neutron scattering experiments \cite{stuart}. Among
the different techniques sensitive to the interaction between flux lines and TB,
transport measurements are very popular. The vortex flow properties are observed to be
anisotropic with respect to the TB directions, but since TB act as coherent barriers for
electronic transport even in the normal state \cite{catherine}, it is not a genuine
pinning effect. It is nevertheless quite clear that TB influence also pinning. The
critical current $I_{c}$ can be measured as a function of the angle $\theta _{B}$ between
the $\overrightarrow{c}$ axis of the crystal and the applied field $\overrightarrow{B}$.
In most of the superconductors, generally untwinned, $\theta _{B}=0$ is a position where
$I_{c}$ is minimum. On the other hand, a maximum is observed in twinned YBaCuO samples,
for moderate magnetic field and temperature values \cite{sanfilippo}. This pinning effect
has been described by different authors \cite{blatter} \cite{feinberg} \cite{sonin}. Even
if the FLL elasticity models used to interpret the data are different, the main idea is
that the direction of TB is a direction along which the flux lines gain energy. The exact
calculation of this energy gain is still an open and difficult question. The reason is
that it depends on the detail of the TB structure, as their concentration in oxygen
vacancies for example. Another possible explanation, compatible with experimental data
\cite{goupil}\cite{maggio}, is to assume that the TB are acting as internal surfaces and
that the pinning effect is merely like a surface screening effect. Some of the
theoretical predictions are the following: For low $\theta _{B}$, it can be more
favorable for flux lines to move away from the direction of $\overrightarrow{B}$ and to
stay along the TB. This corresponds to a so called "locked state". Increasing again
$\theta _{B}$ causes a meandering of the flux lines around the main direction of
$\overrightarrow{B}$. A clear experimental confirmation of such effects necessitates to
observe these FLL deformations using visualization techniques. For instance, Scanning
Tunneling Microscopy have shown a clear alignment of the vortices in rows along the TB
\cite{maggio}. Different studies at low fields were also published, using magneto-optic
and Bitter decoration techniques \cite{gammel} \cite{dolan} \cite{herbsommer}. It was
shown that the magnetic flux penetrates  more easily along the TB than perpendicular to
them, and some information on the symmetry of the flux lines emerging from the surface
has been evidenced. A limit of these experiments is that they do not give information
about the FLL order in the bulk of the samples.

To obtain complementary information and to reveal distortions effects possibly caused by
the TB, it seems highly desirable to use a method that study the FLL structures within
the sample interior itself. Small Angle Neutron Scattering (SANS) provides this
possibility and was used by few groups in twinned YBaCuO \cite{forgan} \cite{keimer}
\cite{delamarre}.

The neutron diffraction pattern presents a fourfold symmetry when the field is applied
parallel to the c-axis. The first interpretation was that this peculiar structure would
arrive as a consequence of a $d_{x^{2}-y^{2}}$ wave function \cite{keimer}. It was
proposed that the combined effects of the $(\overrightarrow{a}, \overrightarrow{b})$
planes anisotropy and of the orientation due to the TB provides a more natural
explanation \cite{forgan} \cite{commentted} \cite{walker}. This second approach was
confirmed by more recent measurements in quasi-untwinned crystals \cite{stuart}. Another
result was that the directions of the flux lines are influenced by the TB directions even
when the field is tilted from the c-axis, with eventually a splitting into two lattices
\cite{stuart}. Despite this good deal of progress and because these experiments were more
devoted to the understanding of the intrinsic symmetry of the FLL, there is still a lack
of detailed knowledge of the TB influence on the FLL.

We present in this paper a SANS study of the interaction between the TB and the FLL. We
have investigated several crystals of YBaCuO possessing different TB densities. Due to
the large magnetic penetration depths in the cuprates (about two or three times that of
pure Niobium for example), the scattering due to the FFL is very small. On the contrary,
the small angle diffusion signal due to the TB themselves exhibits much higher intensity,
that leads to a highly unfavourable ratio FLL signal over background noise. That makes
the overall experiment times consuming, and "only" three single crystals have been
investigated. This study reveals nevertheless that the FLL adopts some systematic
behavior.

\section{Experimental}

In YBaCuO, TB are symmetry planes formed at the tetragonal to orthorhombic transition
along the $<110>$ and $<1\overline{1}0>$ directions, the symmetry operation being a
mirror plane that interchanges crystallographic $\overrightarrow{a}$ and
$\overrightarrow{b}$ axis. We have studied three samples prepared either by Top Seeding
Growth (TSG) or Melt Texturing Growth (MTG). They were chosen because of their large
size, required for such SANS experiments ($m\geq $ few g). Two TSG samples are coming
from MATFORMAG \cite{chaud} in Grenoble (sample I and III), and the MTG sample is coming
from the CRISMAT Laboratory in Caen (sample II) \cite{delamarre}. The T$_{c}$ were about
92K after a week of annealing under pure oxygen. The samples were studied by high
resolution electron microscopy to determine the average distance between TB. The results
were 25 nm for sample I, 50 nm for sample II and 57 nm  for sample III (Fig. 1). The size
of the twinned domains is large (few microns). However, in a macroscopic sample, both orientations $(110)$ and $(1\overline{%
1}0)$ of the planes are observed. The nature itself of the TB can be also very different
according to the sample preparation, ranging from very clean and thin TB to very dirty
ones of few nm wide with a lot of punctual defects and oxygen vacancies.
 The SANS measurements were carried out using the PAXY instrument at the LLB (Saclay, France). The
magnetic field was 0.2T or 0.5T. The neutron wavelength was $\lambda _{n}=10$%
\AA\ or $\lambda _{n}=15$\AA\ ($\delta \lambda _{n} /\lambda _{n}\approx 10$ \%)
depending on the spacing of the Bragg planes formed by the flux lines. The incident beam
divergence was about $0.15{{}^{\circ }}$. As previously noted \cite{forgan}, the Bragg
peaks in such samples were found large enough to fulfill Bragg conditions and to
illuminate the 2D detector without performing complete $\omega$ rocking curves. Let us
define $\theta _{B}$ as the angle between the applied magnetic field and the
crystallographic c-axis, and $\psi$ as the angle between the magnetic field and the
c-axis (Fig. 2). The rotation axis is $(110)$.

 For the SANS study, the
neutron beam was first aligned with the magnetic field and the c-axis with an accuracy of
about 0.1 degree, by observing the (strongly anisotropic) diffusion along the TB and
perpendicular to the c-axis in the normal state. The scattered neutrons were collected
by a XY multidetector (128 $\times $ 128 cells of 0.5 $\times $ 0.5 cm$^{2}$%
) located at a distance of 7 m from the sample. The diffraction patterns were taken at
4.2 K, after a field cooling process. The background was recorded for all patterns at 100
K in the normal state, with the magnetic field inside the sample. In YBaCuO at a low
temperature, the critical current is strong enough to keep the applied magnetic field in
the sample even after its removal. Some patterns were recorded in this remanent mode, for
experimental convenience. We observed no difference between the data recorded following
this method or with the magnetic field applied during the measurement.

\section{Results and discussion.}

\subsection{ SANS patterns: general points}

For the three samples, the FLL diffraction patterns always exhibit distinct Bragg peaks,
characteristics of long range order. Twin boundaries are extended defects. High
resolution microscopy allows us to know that they are spaced by an average value $\langle
d \rangle$, with a broad distribution around this value ($\frac{\Delta d}{\langle d
\rangle}\approx 80\%$ for the sample III, as shown in Fig 1). Nevertheless, the position
of the observed Bragg peaks is fixed in the reciprocal space by Q=$\frac{2\pi}{a_{0}}$
($a_{0}$ is the spacing between flux lines given by the applied magnetic field value).
This was observed even for a FLL that is dilute compared to the TB density. Moreover,
contrary to the diffraction ring observed in polycrystalline samples, the peaks are
resolved in the azimuthal direction (orientational order).
 The long range order of the vortices appears strong enough to resist to the
localization that such numerous and extended defects could induce. The analogy with a
glassy state, such as \textquotedblright Vortex Glass\textquotedblright\ or
\textquotedblright Bose Glass\textquotedblright\ whom signature would better be a ring of
scattering, appears thus not so adequate here.

\subsection{The interaction between the FLL and the TB}

Let us first discuss the measurements made on the sample III for which the main spacing
between TB is 57 nm (Fig. 1). The applied magnetic field is 0.5 T, so as to reach a flux
lines density nearly equal to that corresponding to the TB density. Fig. 3 exhibits
typical diffraction patterns obtained after the background subtraction, in the
configuration $\theta _{B}=\psi$ and for various angles $\theta _{B}$. For large enough
angles $\theta _{B}\gtrsim$ 10 deg, the diffraction pattern looks like a classical
pattern given by an hexagonal lattice, but with a distortion imposed by the anisotropy
between the $(\overrightarrow{a},\overrightarrow{b})$ plane and the
$\overrightarrow{c}$-axis. One can notice that the two vertical spots are more intense.
As the scattered intensity scales as $\frac{1}{\lambda^{4}}$ with $\lambda$ the London
length, all Bragg peaks are not of equal intensity in an anisotropic superconductor. The
vertical Bragg peaks are intrinsically more intense because they correspond to the
direction where $\lambda$ is small. This effect can be reinforced by the good alignment
of the flux lines along the $(110)$ TB family. For $\theta _{B}=0$ deg, the pattern is
very peculiar, as already pointed out by several authors
\cite{keimer}\cite{forgan}\cite{commentted}\cite{walker}. In addition to the two vertical
spots, two horizontal spots of strong intensity are obtained. These four spots are
arranged along the two TB directions $(1\overline{1}0)$ and $(110)$: it is thus natural
to assume that they reveal the alignment of vortices along the TB and that vertical and
horizontal directions are equivalent by symmetry. In addition, it appears that some
spots, at an apex angle $\beta \approx 47(\pm 2)deg$ from the vertical ones, are less
defined. This observation was first made in \cite{forgan} and \cite{keimer}. In order to
interpret this value, Forgan et al. have proposed to take into account both the
orientational effect of the TB and the in-plane anisotropy of the magnetic penetration
depths \cite{commentted}. The peculiar value of the apex angle can be attributed to an
(a,b) anisotropy of $\frac{\lambda _{a}}{\lambda _{b}}\approx 1.3$ \cite{walker}, which
is indeed observed for good oxygenated crystal \cite{wang}. Following the same idea, we
can also try to interpret the present pattern. The (a-b) anisotropy imposes a monoclinic
unit cell. Due to the symmetry operation of the two TB families, there is a sum of four
monoclinic cells differently oriented (two orientations in each twinned domain). There is
nevertheless a notable difference between what we observe and the patterns observed in
\cite{keimer} and \cite{forgan}. This is evidenced in the first pattern of Fig. 3 where
one spot of each domain is apparently missing. As the orientation of one unit cell is not
intrinsic but rather due to its interaction with the TB, we think that the magnetic form
factor of the vortices can be also modified in the same way that it is modified by an
image effect due to a non superconducting surface (the Bean-Livingston barrier
\cite{bean}). As a consequence of this reasonable assumption, the supercurrent
distribution surrounding the vortices is perturbed by the TB and the barrier effect
stretches out the supercurrent loop in the direction of the TB (fig.4). It explains why
the intensity of the spot named $q_{1}$ in the fig.4 is very low and that this spot is
not observed. We emphasize that, contrary to \cite{forgan} and \cite{keimer} who worked
at much higher magnetic field, our experiments employed a density of flux lines
comparable to the TB density. Such boundary-like effects are less expected at high field
where supercurrents largely overlap.

In the case of a conventional FLL with straight flux lines, the reciprocal plane
perpendicular to the magnetic field is the only one that contains Bragg peaks. In twinned
YBaCuO, diffraction peaks appears also out of this plane as soon as $\theta _{B}$ differs
from zero \cite{forgan}\cite{delamarre}. This plane is the one perpendicular to the c
axis of the crystal (fig.5) and exhibits spots that are along the direction of the (110)
TB family. They are very intense, comparable to the vertical ones. In order to explain
the existence of these additional peaks, a possibility is that the FLL is locked along
the direction of the TB (the locking phase) and does not follow the applied magnetic
field anymore \cite{blatter} \cite{zhukov}. But in this case, any Bragg peak is expected
in the plane perpendicular to the magnetic field contrarily to what is observed in Fig.
3. It was also proposed that the extra peaks correspond to the diffraction pattern of two
FLLs pinned by the TB, while another FLL follows the magnetic field to give the distorted
hexagonal pattern presented in fig.3 \cite{forgan}. In such a case, the peaks observed in
the two different planes are coming from two different unit cells and a special
relationship between them is not expected \cite{delamarre}. As we will describe now, we
rather observed a clear relationship between the two families of Bragg peaks and a model
treating these peaks as coming from the same unit cell fit well with the data.

In fig. 6, we have reported a schematic drawing of the reciprocal space and defined the
peaks, $\overrightarrow{q_{2}}$ for the vertical common to both planes,
$\overrightarrow{q_{1}}$ and $\overrightarrow{q_{3}}$ for the others in the plane
perpendicular to the magnetic field, and $\overrightarrow{m*}$ in the plane perpendicular
to the c axis). The angle $\beta$ is the angle between $\overrightarrow{q_{2}}$ and
$\overrightarrow{q_{1}}$ in the reciprocal space. Fig. 7 is a schematic drawing of the
equivalent real space. In the plane perpendicular to the magnetic field B, the hexagonal
lattice distorted by the anisotropy is represented by a unit cell with the vectors
$\overrightarrow{a_{1}}$ and $\overrightarrow{a_{2}}$. They correspond to the vectors
$\overrightarrow{q_{1}}$ and $\overrightarrow{q_{2}}$ in the reciprocal space. The third
vector of interest is
$\overrightarrow{q_{3}}=\overrightarrow{q_{1}}-\overrightarrow{q_{2}}$, and is hence a
vector of the lattice. In order to interpret the presence of the spots located at $\pm
\overrightarrow{m*}$, we have assumed that the vortices meander along their axis to keep
a part aligned with the TB, as suggested in the theoretical models described in the
introduction of the paper. Let us call $\overrightarrow{p}$ the vector representing the
periodicity of this modulation along the direction of B. A direct observation of
$\overrightarrow{p*}$ (perpendicular to B) is really difficult because of the non
magnetic scattering which is very intense in this direction. Nevertheless, one can expect
to observe a combination of $\overrightarrow{q_{1}}$, $\overrightarrow{q_{2}}$ and
$\overrightarrow{p*}$. The diffraction spots located at $\pm \overrightarrow{m*}$ indeed
correspond to one of such combination:
$\overrightarrow{m*}$=$\overrightarrow{p*}$+2$\overrightarrow{q_{1}}$-$\overrightarrow{q_{2}}$.
We have also to understand why these peaks are really intense (even certainly the most
intense) and are those observed.  A natural explanation is obtained if one considers the
Fourier-transform of the flux lines form factor. As $\lambda_{ab} < \lambda_{c}$ and as
the diffracted intensity scales as $\lambda ^{-4}$, the form factor "shape" looks like a
plane containing the c-axis and perpendicular to the
($\overrightarrow{B}$,$\overrightarrow{c}$) plane. The Fourier transform of this plane is
a single line being situated in the $(\overrightarrow{B},\overrightarrow{c})$ plane and
perpendicular to the $\overrightarrow{c}$-axis. Then, the main part of the scattered
intensity is along this line, that indeed contains $\overrightarrow{m*}$.

This simple model of meandering vortices allows to explain qualitatively the existence of
the Bragg peaks which are observed on the two different planes of the reciprocal space.
They are moreover some quantitative predictions which can be tested. First, in the case
of a single FFL which can be described with a unique unit cell, the angle $\beta$ can be
related to $\theta_{B}$ using Kogan$^{^{\prime}}$s scaling law \cite{kogan}:

\begin{equation}
\tan \beta =\frac{\sqrt{3}}{R^{2}},\text{ \ }R=\sqrt{cos^{2}\theta _{B}+\gamma
^{-2}sin^{2}\theta _{B}}
\end{equation}

In fig.8, we have reported the variation of $\beta $ as a function of $\theta_{B}$ and
compared it to this equation. The agreement is very good down to a critical angle $\theta
_{Bcri}$ of about 10 degrees. The electronic anisotropy extracted from the fit is $\gamma
=4.7\pm 1.5$ for the sample III, which is a little smaller than the most common values
(5-10) obtained by transport or torque experiments but in good agreement with previous
measurements using SANS \cite{forgan}. But one has to note that compared to indirect
methods, SANS directly probes the anisotropy of the bulk London lengths.

The model of the flux lines meandering \cite{delamarre}, using simple geometrical
arguments, makes a prediction on the ratio $\overrightarrow{m*}/\overrightarrow{q_{2}}$,

\begin{equation}
\frac{m*}{q_{2}}=\frac{a_{2}sin\beta
cos\theta_{B}}{a_{0}}=\frac{\frac{\sqrt{3}}{2}}{\sqrt{1+{{(\frac{tan\theta _{B}}{\gamma
})}^{2}}}}
\end{equation}

The agreement between the experimental data and this model is promising (see fig.8),
again down to an angle of about $\theta _{Bcri} \approx 10 deg$. One can notice that this
agreement extends at least up to $\theta _{B}$ = 60 deg, which could signify that the
meandering of vortices is present even for very large angles $\theta _{B}$. If one uses
previous data obtained by Delamare et al. \cite{delamarre}, who made same kind of
measurements on two different YBaCuO samples, we can apply the same model and we obtain
different critical angles $\theta _{Bcri}$ (Fig. 9). For the sample I, $\theta _{Bcri}$
is more than 60 degrees and about 40 degrees for the sample II. Keeping in mind that the
main spacing between TB was 25 nm in sample I and 50 nm in sample II, this shows that the
value of the critical angle increases when the TB density decreases. It is worth noting
that the TB density is not the unique structural difference between the three samples.
Electron microscopy studies reveal different types of defects in the samples, such as
$BaCeO_{3}$ inclusions and non superconducting "green phase" ($Y_{2}BaCuO_{5}$) of few
microns size in the MTG sample (sample II) that are not present in the samples I and III.
Beyond the presence of those different defects, the direction and the distance between TB
seem to be the relevant parameter that controls the FLL structure in such YBaCuO samples,
at least as far as the resolution of a SANS experiment is involved. Extended defects,
which can act as internal surfaces or barriers for the vortices, are indeed much more
efficient than smaller defects particles for perturbing the FLL structure. To change the
ratio of the flux lines density over the TB density, it is also possible to change the
value of the magnetic field. We have chosen a value of 0.2T to increase the distance
between the flux lines. A comparison of the variations of $\beta$ as a function of
$\theta _{B}$ for sample III is reported in fig.10. The critical angle is higher than for
a magnetic field of 0.5T and is above 35 degrees. It shows again that the critical angle
is larger when the ratio of the flux lines density over the TB density is lower.

In conclusion of this part and as summarized in fig. 11, $\theta_{B}>\theta_{Bcri}$
corresponds to a state where the FLL can be described with a unique anisotropic unit
cell, but where the flux lines are also meandering around their mean direction. Such
analysis breaks down for $\theta_{B}<\theta_{Bcri}$, with $\theta_{Bcri}$ determined by
the ratio of the flux lines density over the TB density.

\subsection{The distorted state}

Let us now discuss the nature of the FLL state which is observed for $0<\theta _{B}\leq
\theta _{Bcri}$. First and surprisingly, looking carefully at the patterns obtained with
the neutron beam along the magnetic field, it is observed that the relationship
$\overrightarrow{q_{1}}=\overrightarrow{q_{2}}+\overrightarrow{q_{3}}$ does not hold
anymore. It follows that all the spots do not belong to the same cell. Another sign of a
change in the FFL structure can also be evidenced on fig.8 by the simple fact that
$\beta$ does not follow the equations (1) anymore, but takes a rather constant value of
47 degrees from $\theta _{B}=\theta _{Bcri}$ down to $\theta _{B}=0$. Now if one
introduces this value in the calculation of $m*/q_{2}$ (equation (2)), the agreement with
the experimental data is rather good (see fig. 9). This test is important because it
shows that we can continue to apply the model of flux lines meandering, even if a
description using a unique unit cell is no more appropriate. We note that the value of
$\beta$ is not changing when $\theta_{B}$ decreases. It is likely that the interpretation
proposed when $\theta_{B}$ is strictly zero is still valid. A fourfold symmetry is
observed when $\theta_{B}$=0, that was previously interpreted as coming from the sum of
four monoclinic domains with one peak turned off in each cell. As soon as $\theta_{B}$
moves away from zero, the degeneracy between the two families of FFL breaks down. This is
because the (110) TB family is not perpendicular to the diffraction plane anymore and
that one symmetry element has vanished. We observe then a sum a two monoclinic domains,
with one spot missing in each domain. This leads to the pattern as we observe.

Finally, let us discuss a possible consequence of this FLL deformation for the value of
the critical current. It is usually observed that heavily twinned samples display
stronger critical currents than untwinned ones for an angle
$(\overrightarrow{B},\overrightarrow{c})<\theta _{cr}$ and at low field. The breakdown of
the collective pinning regime and the crossover to strong pinning of single vortices in
the twin boundaries can be involved. Another possibility is that TB are interfaces which
are favouring the development of shielding currents. The observed distortion of the FLL
for small $\theta _{B}$ angles, that we interpret with the deformation of the local
magnetic field distribution, looks more consistent with this second interpretation. This
is also in agreement with the very high currents measured along the TB by Maggio et al
\cite{maggio}. Following this idea, the TB have to be considered as normal planes along
which flows a surface-like current. At the scale of the sample, this can correspond to an
important critical current. It has been shown that the strong FLL distortions induced by
the boundaries are screened in a healing length $\lambda _{v}$ \cite{sonin}, coming from
the non-local elasticity of the flux lines array. For not too large angles, its order of
magnitude is given by $a_{o}$ perpendicular to the flux lines \cite{goupil}, i.e.
$a_{o}.cos \theta$ perpendicular to the TB. When $a_{o}.cos \theta$ approaches $<d>$, the
FLL distortions are affecting the whole FLL. One can thus take as a first approximation
and as a criteria $a_{o}.cos \theta \gtrsim <d>$ for the beginning of strong FLL
deformation. This simple expression approximates reasonably well the experimental data
(Fig. 11).

It is worth noting that such distortions effects are observed at low fields in heavily
twinned samples. They are not to be confused with FLL symmetry change, from triangular to
square lattice, which have been very recently observed in YBaCuO at high fields and
possibly coming from d-wave character of the order parameter \cite{brown}.

\section{Conclusion}

In conclusion, a neutron diffraction study of the effect of the twin boundaries density
on the Flux Lines Lattice in YBaCuO has been presented. We observe that the FFL moves
away from its ideal behavior in two ways. Flux lines meander around the main direction of
the applied magnetic field to keep a part along the TB. The resulting kinks display a
long range order, despite the disorder of the TB at the scale of the sample. Furthermore,
two different FLL structures are observed according to the angle $\theta _{B}$ between
the magnetic field and the $\overrightarrow{c}$-axis, depending on the ratio of the TB
density over the flux lines density. At small angles, the FLL structure is distorted.
This is interpreted with an elongated surrounding supercurrent around the flux lines, due
to the border effect of the TB. A cross-over at large angle or at small Flux Lines
density towards the more usual anisotropic structure is observed.

Acknowledgments: The authors would like to thank different members of the CRISMAT
Laboratory: Maryvonne Hervieu for the electron microscopy studies of the YBaCuO samples
twinning, Olivier P$\acute{e}$rez and Henry Leligny for helpful discussions. A.P would
like to thank Sylvie H$\acute{e}$bert for her highly appreciated contribution.

\newpage

Figure 1: Determination of the distance between the twin planes from high resolution
electron microscopy in the YBaCuO sample labeled III in this article. The number of twin
planes counted on the picture is plotted versus the domain size.

Figure 2: The orientation of the twin boundaries with respect to the axis of the crystal
and the definition of the different angles used in the SANS experiment.

Figure 3: Typical diffraction patterns obtained at B=0.5T and T=4.2K for the neutron beam
parallel to B ($\psi $ = $\theta _{B}$) at various angle $\theta _{B}$. On the $\theta
_{B}=0$ and $\theta _{B}=60$ deg patterns, the in plane unit cell of the FLL is also
shown. Note that the $\overrightarrow{q_{1}}$ diffraction peak is missing for $\theta
_{B}=0$.

Figure 4: A possible construction of the unit cell for $\psi$ = $\theta _{B}$ $=$ 0.
Because of the deformation of the current around the flux lines, the
$\overrightarrow{q_{1}}$ peak has a too low intensity to be observed. The symmetry due to
the two TB families leads to a total of four domains, as observed in the first pattern of
the fig 3.

Figure 5: Typical diffraction patterns obtained for B=0.5T and T=4.2K for the neutron
beam parallel to the $\overrightarrow{c}$-axis ($\psi$ = 0) at various angle $\theta
_{B}$.

Figure 6: The schematic diffraction patterns obtained with the neutron beam parallel to
$\overrightarrow{B}$ and parallel to the $\overrightarrow{c}$-axis at a given angle
$\theta_{B}$.

Figure 7: A schematic drawing of the real space showing a plane perpendicular to the
magnetic field and a plane containing $\overrightarrow{B}$ and $\overrightarrow{m}$, in
the case of the FLL symmetric state.

Figure 8: The $\theta _{B}$ dependence of the angle $\beta$ and the ratio
$m^{\ast}/q_{2}$ for B=0.5T and T=4.2K. The solid lines are the fits using the model of
flux lines meandering in the case of the symmetric state (equation (1) and (2)). Note the
cross-over at $\theta _{B}$ = $ \theta _{Bcri}\approx 10$ deg.

Figure 9: The $\theta _{B}$ dependence of the angle $\beta $ and the ratio
$m^{\ast}/q_{2}$ (B=0.5T and T=4.2K) for the different YBaCuO samples. The solid lines
are fits using equations (1) and (2) for the "symmetric" state. The dashed line is the
fit putting $\beta \approx 47$ deg in the equation (2)(the "non symmetric" state). Note
the increase of the critical angle for the samples with the highest TB density.

Figure 10: Comparison between the $\theta _{B}$ dependence of the angle $ \beta $ for
B=0.5T and B=0.2T (sample III), showing that the critical angle decreases when the Flux
Lines density (i.e. the magnetic field) decreases.

Figure 11 : Variation of cos($\theta _{Bcri})$ as function of the ratio $a_{o}/d_{twin}$.
The solid line is a linear fit $cos(\theta _{Bcri})= 1.18 (d_{twin}/a_{o})(\pm 0.10)$,
that is close to the criteria explained in the text
$cos(\theta_{Bcri})=(d_{twin}/a_{o})$. The dashed part of the graph corresponds to the
deformed FLL

\end{document}